\documentclass[a4paper]{jpconf}
\usepackage[pdftex]{graphicx}
\usepackage{amsmath}
\usepackage{mathtools}
\usepackage{float}
\usepackage{bm}
\begin{document}
\title{Anomalous dispersion relations in the staggered flux state}

\author{Hiroki Tsuchiura$^{1,2}$, Masao Ogata$^{3}$, Kunito Yamazaki$^{1}$, and Rui Asaoka$^{1,2}$}

\address{
$^1$Department of Applied Physics, Tohoku University, 6-6-05 Aoba, Sendai 980-8579, Japan\\
$^2$Center for Spintronics Research Network, Tohoku University, Sendai 980-8577, Japan \\
$^3$Department of Physics, University of Tokyo, Bunkyo, Tokyo 113-0033, Japan
}
\ead{tsuchi@solid.apph.tohoku.ac.jp}

\begin{abstract}
We study the quasiparticle properties in the d-wave superconductivity, antiferromagnetism,
 and the staggered flux state within a renormalized mean-field theory based on the two-dimensional $t$-$J$ model.
In particular, we focus on the anomalous quasiparticle dispersion relations around the antinodal region
of the Brillouin zone argued in Hashimoto {\it{et al.}}, Nature Phys. {\bf 6}, 414 (2010).
We obtain the qualitatively consistent results with their observations when we take account of 
the SF state.
The present analysis shows that the SF order can be a possible candidate of symmetry-breaking
pseudogap states coexisting with the dSC.
\end{abstract}

\section{Introduction}
Recent experimental studies have indicated that the pseudogap phase in underdoped high-Tc cuprates
is accompanied by some phase transitions with the breaking of time-reversal \cite{Kaminski,Li}, 
rotational \cite{fujita_rot}, or translational symmetry. 
These symmetry breakings imply that there can be nematic order, charge order, or other instabilities 
in the pseudogap phase\cite{cdw1,fold1,fold2,fold3,fold4}. 
It has also been suggested that the pseudogap order and d-wave superconductivity (dSC) can coexist 
within the superconducting dome of the cuprates at $p < 0.19$, where p is the hole concentration.

One of the intriguing experimental evidences of the coexisting pseudogap phase with dSC
is a gap-structure in the AREPES data opening with broken particle-hole 
symmetry found in Bi-2201 \cite{hashimoto1,hashimoto2}.
They examined the quasi-particle dispersions along the Brillouin zone boundary,
and claimed that the gap opens below the $T_{c}$ is distinct from dSC
based on a phenomenological model calculation, and also that the gap can be
understood by assuming the coexisting antiferromagnetic (AF) order with dSC.
Soon after the that, Greco and Bejas \cite{greco} have carried out theoretical calculations
of the quasiparticle properties in the pseudogap state incorporating the effects
of $d$ charge-density-wave fluctuation, or the so-called staggered flux fluctuation) 
based on the $t$-$J$ model.
They successfully described the anomalous dispersion relations along the zone-boundary
without taking account of AF order or fluctuation.

Motivated by these backgrounds, we here study the quasiparticle properties 
in dSC, AF, and also the staggered flux (SF) states within the same theoretical framework.
The SF state is a normal, or non-superconducting state with broken symmetries, and
has been studied by many groups since the early stage of research on cuprate
superconductors \cite{affleck,lee,Laughlin,morr,zhou,yokoyama}.
It was considered firstly as a candidate of the ground state for the CuO$_{2}$ plain, 
then has been studied as a candidate for a non-superconducting state that causes the pseudogap
found in underdoped cuprates.
In this paper, we study the quasiparticle dispersions in the SF, AF, and dSC states by using
the $t$-$J$ model within the Bogoliubov-de Gennes theory based on an extended version of the Gutzwiller
approximation.
We will see the SF state exhibits the qualitatively consistent dispersion relations with those observed in
the ARPES experiments \cite{hashimoto1,hashimoto2}.
%

\section{model}
In this study, we employ an advanced version of the renormalized mean-field theory
for the $t$-$J$ model, the Hamiltonian of which is given as
\begin{equation}
{\mathcal H} = -\sum_{\langle i,j \rangle,\sigma}
P_{G}( t_{ij}c^{\dag}_{i\sigma} c_{j\sigma} + {\rm h.c.} )P_{G}
 + J\sum_{\langle i,j\rangle}\mbox{\boldmath $S_{i}\cdot S_{j}$}
 - \mu\sum_{i,\sigma}c^{\dag}_{i\sigma} c_{i\sigma}
\label{hamil}
\end{equation}
in the standard notation where $\langle i,j\rangle$ means the summation
over nearest-neighbor pairs.
The Gutzwiller's projection operator $P_{G}$ is defined as
$P_{G} = \Pi_{i}( 1 - n_{i\uparrow}n_{i\downarrow})$.
%
The Bogoliubov-de Gennes (BdG) equation for a renormalized mean-field Hamiltonian based on
an extended Gutzwiller approximation \cite{ogata,HTmagimp,HTvortex} is given as
\begin{equation}
\left(
\begin{array}{cc}
H_{ij}^{\uparrow} & F_{ij} \\
F_{ji}^{*} & -H_{ji}^{\downarrow}
\end{array}
\right)
\left(
\begin{array}{c}
u_{j}^{\alpha} \\
v_{j}^{\alpha}
\end{array}
\right)
= E^{\alpha}
\left(
\begin{array}{c}
u_{i}^{\alpha} \\
v_{i}^{\alpha}
\end{array}
\right)
,
\end{equation}
with
\begin{eqnarray}
H_{ij}^{\sigma} 
&=& -\sum_{\tau}\left( t_{ij}^{\rm eff} + J_{ij}^{\rm eff}\tilde{\chi}_{ji}
\right) \delta_{j,i+\tau}  
-\sum_{\nu}t_{ij}'^{\mathrm{eff}}\delta_{j,i+\nu}
+ \sigma\delta_{ij}\sum_{\tau}h_{i,i+\tau}^{\rm eff}
 - \mu\delta_{ij}
\nonumber \\
 F_{ij}^{*} &=& -\sum_{\tau} J_{ij}^{\rm eff}\tilde{\Delta}_{ij}
 \delta_{j,i+\tau}  ,
\end{eqnarray}
where $\sigma = \pm 1$,
$i+\tau$ represents the nearest neighbor sites of the site $i$, 
while $i+\nu$ the 2nd neighbor sites.

The renormalized parameters $t_{ij}^{\rm eff}$, $t_{ij}'^{\mathrm{eff}}$, $J_{ij}^{\rm eff}$ and 
$h_{ij}^{\rm eff}$ have somewhat complicated forms depending on the local expectation values 
$\Delta_{ij}=\frac{1}{2} \langle c_{i\uparrow}^{\dag}c_{j\downarrow}^{\dag}
 - c_{i\downarrow}^{\dag}c_{j\uparrow}^{\dag}\rangle$, 
$\chi_{ij} = \langle c_{i\sigma}^{\dag}c_{j\sigma}\rangle$, and
$m_i=\frac{1}{2}\langle n_{i\uparrow}-n_{i\downarrow}\rangle$
\cite{ogata,HTmagimp,HTvortex}.
These parameters are determined from cluster calculations to reproduce the variational Monte Carlo
results \cite{ogata}.
Thus the present renormalized mean-field theory can give more reliable results
than theories based on the conventional slave boson or Gutzwiller approximation \cite{zhang}.
In particular, the overestimation of AF commonly seen in the conventional mean-field theories is improved
in the present theoretical treatment.
Since the detailed derivation of these parameters are given in ref. \cite{ogata}, here we only show
in Table 1 the self-consistently obtained numerical values of them for the $t$-$J$ model with $J/t = 0.3$,
$t'/t = -0.3$, and the hole doping $\delta = 0.09$.
In the present calculation, we regard the 2$\times$2 square lattice as a unit cell to solve
the BdG equation numerically.
%

\begin{center}
\begin{table}[h]
\centering
\caption{Renormalized model parameters for the $t$-$J$ model with $J/t=0.3$, $t'/t=-0.3$, and $\delta = 0.09$.
Here we omit the site indices $i$ and $j$ because we are studying homogeneous electronic states.} 
\begin{tabular}{{cccccccc}}
\br
state & $t^{\mathrm{eff}}$ & $t'_{\mathrm{eff}}$&
	 $J^{\mathrm{eff}} \tilde{\varDelta}$ & $J^{\mathrm{eff}} \tilde{\chi}$ & $h^{\mathrm{eff}}$ & $\mu$  \\
\mr
normal state & 0.1445 & -0.0495 & 0 & 0.0805 & 0  & -0.161\\
dSC & 0.1403 & -0.0494   & 0.032  & 0.0642 & 0   & -0.143\\
SF & 0.1416 & -0.0497   & 0  & 0.067$\pm$0.0332$i$ & 0 & -0.165\\
dSC+AF & 0.1384  & -0.0491  & 0.030   & 0.062 & 0.0285 & -0.144\\
\br
\end{tabular}
\end{table}
\end{center}

\section{Results and discussion}
Let us look at firstly the quasiparticle states of dSC and SF states.
Figure 1 shows the density of states (DOS) per each site for dSC and SF states.
Here the DOS is given as
\begin{eqnarray}
 N_{j}(E) = \frac{1}{N_{c}} \sum_{\bm{k},\alpha}
 	 \left[ \left| u_{j}^{\alpha}({\bm k}) \right|^{2} \delta( E^{\alpha}({\bm k}) - E)
	 + \left| v_{j}^{\alpha}({\bm k}) \right|^{2} \delta( E^{\alpha}({\bm k}) + E)
	  \right]
\end{eqnarray}
where $j$ represents a site where the local density of states is calculated, $\alpha$ is the eigenstate index,
and ${\bm k}$ is the Bloch wave number of the present unit cells. 
The blue line in Fig. 1 represents the DOS for dSC, and the red one for SF states.
The form of the FS implies that the DOS at zero-energy is nonzero, in contrast to the results for dSC,
as previously found by Morr \cite{morr}.
In particular, we can see that the DOS for the SF state is shifted downwards in energy,
in comparison to the case of dSC.
Thus we can expect the quasiparticle excitation energy for the SF is larger than that of dSC
around the antinodal region of the Brillouin zone.
\begin{figure}[htb]
\begin{center}
\includegraphics[clip,width=8cm]{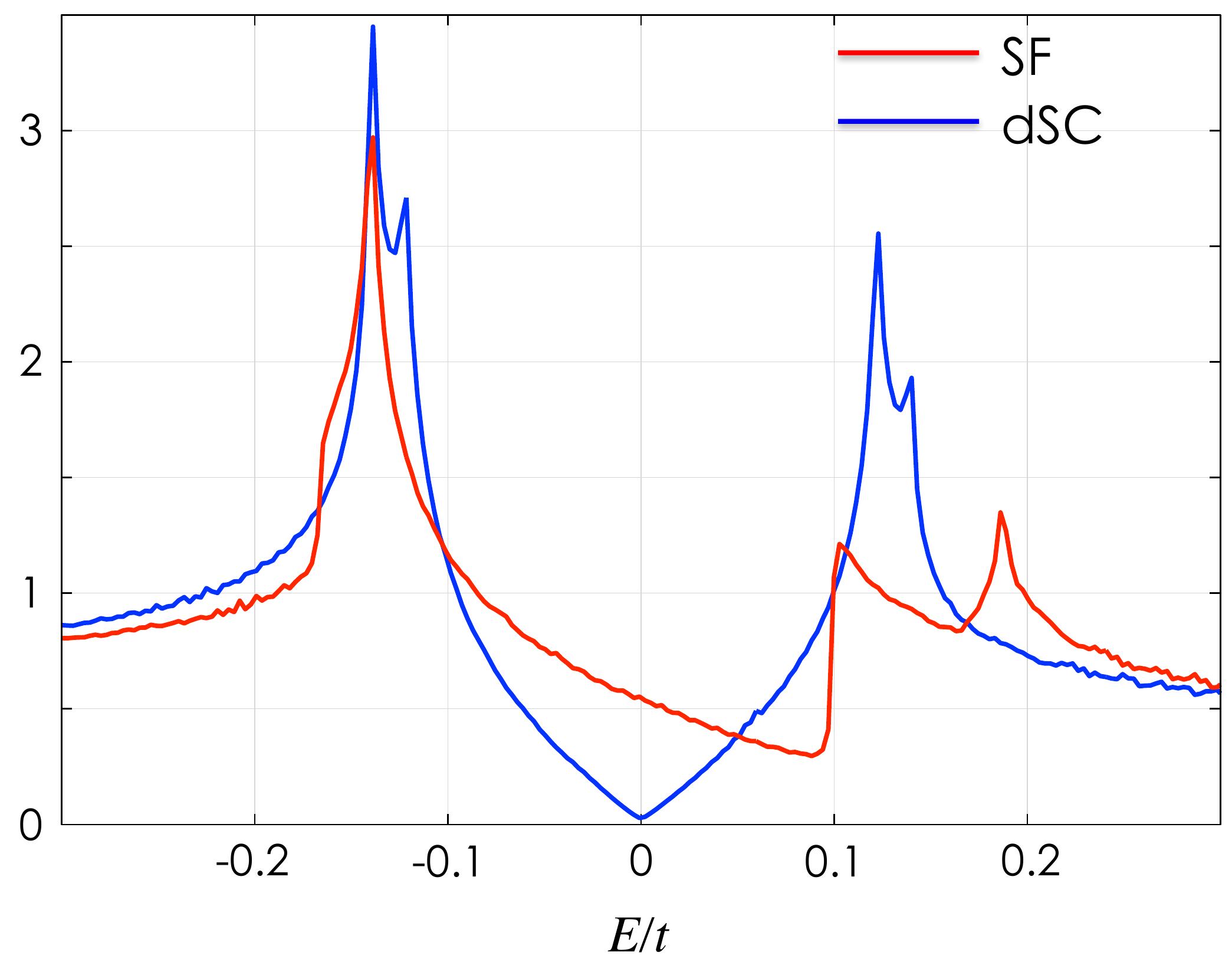}
\end{center}
\caption{
The density of states (DOS) per each site for dSC (blue) and SF (red) states.
}
\label{dSF}
\end{figure}

\begin{figure}[htb]
\begin{center}
\includegraphics[clip,width=8cm]{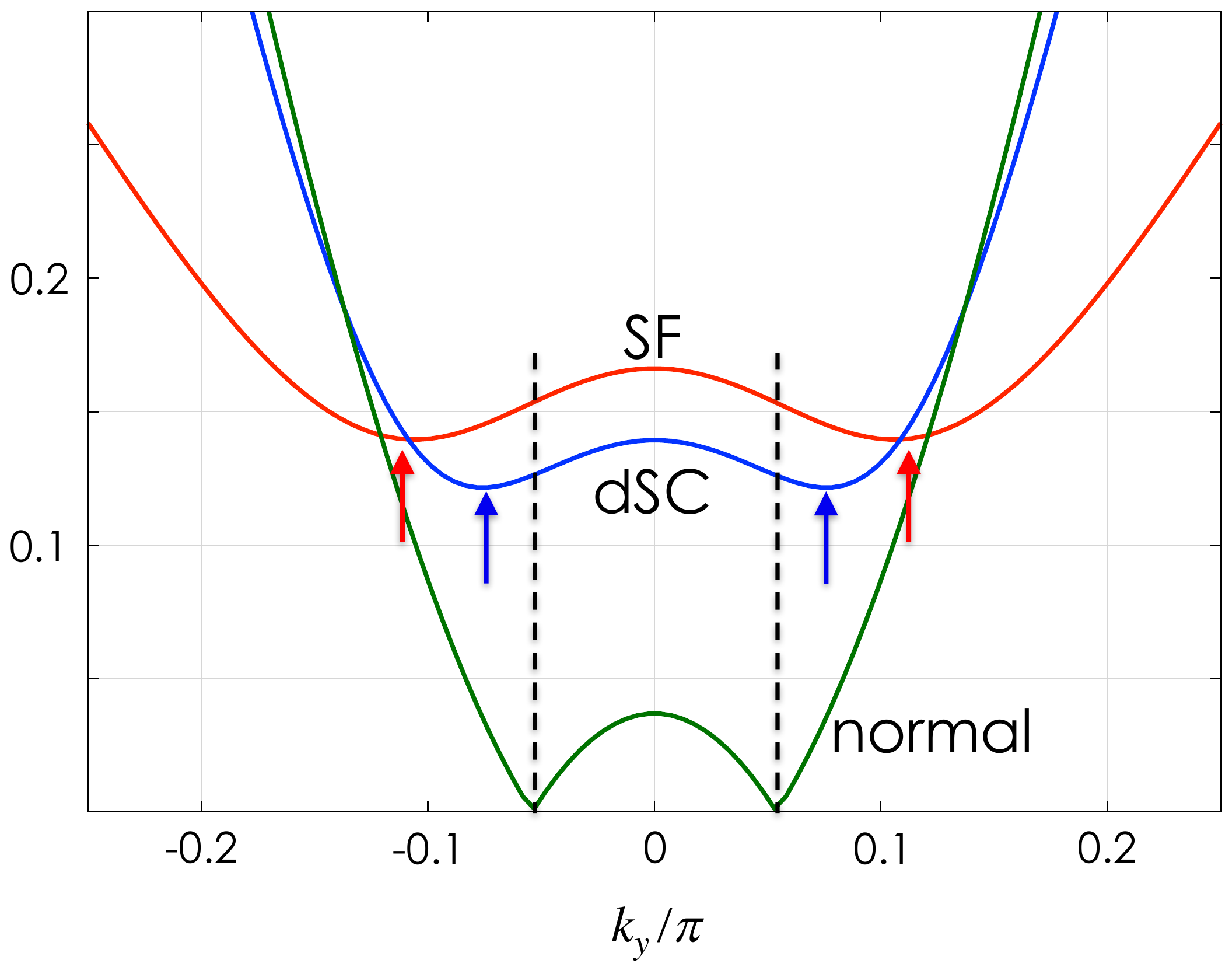}
\end{center}
\caption{
Quasiparticle dispersion relation for dSC (blue), SF (red), and the normal state (green)
along the Brillouin zone boundary $k_{x}=\pi$.
}
\label{disp}
\end{figure}

Next, we show the antinodal quasiparticle dispersions $E_{\mathrm{dSC}}({\bm k})$ for dSC (blue)
and $E_{\mathrm{SF}}({\bm k})$ for SF (red) \cite{SFdisp}
along the Brillouin zone boundary $k_{x} = \pi$ in Fig. \ref{disp}.
We also plot the normal state dispersion relation (green) for comparison.
As expected, the dSC state shows the superconducting energy gap
 $\varDelta E_{\mathrm{dSC}} \sim 0.12$
around $k_{y}/\pi \sim \pm 0.08$ as indicated by blue arrows.
This momenta with the gap minima is slightly different from the Fermi wave number in the normal state
due to the chemical potential shift accompanying the superconducting phase transition, as shown in
Table 1.
The SF state exhibits the larger excitation energy gap $\varDelta E_{\mathrm{SF}} \sim 0.14$ around 
$k_{y}/\pi \sim \pm 1.1$, which is markedly away from those for the dSC state, as indicated by red arrows 
in Fig. \ref{disp}.
Since the chemical potential in the SF state is almost the same as the normal state (Table 1),
the marked difference of the momenta with the gap minima from the Fermi wave number is due to
the broken particle-hole symmetry in the SF state.
These results are consistent with the recent ARPES observations \cite{hashimoto1,hashimoto2}
in which an anomalous dispersion relation has been found at $T<T_{c}$ in an optimally doped Bi-2201
cuprate superconductor.
Thus the SF state can be a possible candidate of symmetry-breaking pseudogap states coexisting with
the dSC.
We note here that, in the present calculation, the SF state does not have a lower energy than the dSC,
and also that the coexisting state of these two (dSC+SF) has not been found.
We can expect, however, that the SF or the SF+dSC state may be stabilized when we consider
finite temperature effects, or intersite Coulomb repulsion \cite{zhou}.
%

\begin{figure}[htb]
\begin{center}
\includegraphics[clip,width=14cm]{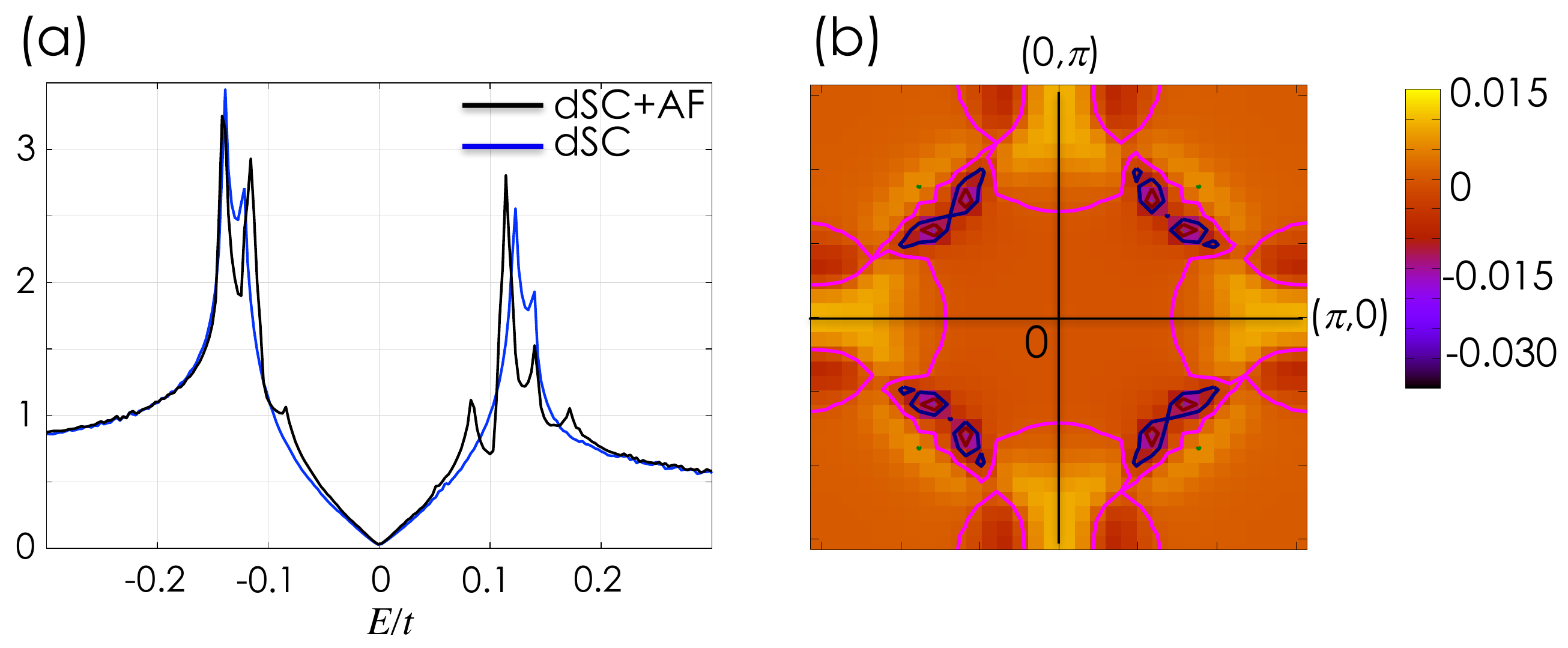}
\end{center}
\caption{
(a) The DOS for the dSC state coexisting with AF order.
(b) The difference of the momentum distribution functions between the dSC and the coexisting state.
}
\label{dAF}
\end{figure}

Finally, let us mention the effects of antiferromagnetic (AF) order on the anomalous dispersion relation
found in the zone boundary shown in Fig. \ref{disp}.
If the AF order is taken into account in our calculation, we find a small expectation value of
$\langle S_{j}^{z} \rangle \sim \pm0.05$ coexisting with the dSC state for the hole doping $\delta=0.09$.
Thus the antiferomagnetic gap is also very small, and we confirm that the DOS for this coexisting state
is almost similar with that of the dSC, as shown in Fig. \ref{dAF} (a). 
A small gap-like structure can be found around $E/t \sim 0.1$.
In order to clarify the effect of this small gap structure in the DOS, 
we calculate the momentum distribution function $n({\bm k})$ for the coexisting state (dSC+AF) and
the dSC state. 
Figure \ref{dAF} (b) shows the difference of the $n({\bm k})$ between these two states 
in the full Brillouin zone.
We can clearly see that the effects of the coexisting AF order is concentrated around the AF hot spots,
and does not have a significant effect around the antinodal region of the Brillouin zone.
Thus we speculate that the anomalous dispersion relation found in the ARPES experiments 
\cite{hashimoto1,hashimoto2} is not due to AF.

\section{Summary}
We have studied the quasiparticle dispersion relation around the antinodal region of the
full Brillouin zone  in the dSC and SF states within the renormalized mean-field theory
based on the $t$-$J$ model.
The present analysis has shown that the SF order can be a possible candidate of symmetry-breaking
pseudogap states coexisting with the dSC.

\section{Acknowledgments}
This work was supported by JSPS KAKENHI Grant Number JP17K05528.
The numerical computations were carried out at the Cyber-science Center, Tohoku University.


\begin{thebibliography}{9}
\bibitem{SFdisp}
 A detailed derivation of the dispersion relation $E_{\mathrm{SF}}({\bm k})$ is given, for example, 
 in the Appendix A of \cite{yokoyama}.
\end{thebibliography}
\end{document}